# ON THE COMPUTATIONAL MODELING OF THE VISCOSITY OF COLLOIDAL DISPERSIONS AND ITS RELATION WITH BASIC MOLECULAR INTERACTIONS

A. Gama Goicochea[1, 2], M.A. Balderas Altamirano[1], R. Lopez-Esparza[1, 3*], M. A. Waldo[2],

E. Perez[1]

[1]Instituto de Física, Universidad Autónoma de San Luis Potosí, San Luis Potosí, Mexico.

[2]A. Schulman of México, San Luis Potosí, Mexico.

[3]Departamento de Física, Universidad de Sonora, Hermosillo, Sonora, Mexico

## ABSTRACT

The connection between fundamental interactions acting in molecules in a fluid and macroscopically measured properties, such as the viscosity between colloidal particles coated with polymers, is studied here. The role that hydrodynamic and Brownian forces play in colloidal dispersions is also discussed. It is argued that many – body systems in which all these interactions take place can be accurately solved using computational simulation tools. One of those modern tools is the technique known as dissipative particle dynamics, which incorporates Brownian and hydrodynamic forces, as well as basic conservative interactions. A case study is reported, as an example of the applications of this technique, which consists of the prediction of the viscosity and friction between two opposing parallel surfaces covered with polymer chains, under the influence of a steady flow. This work is intended to serve as an introduction to the subject of colloidal dispersions and computer simulations, for last – year undergraduate students and beginning graduate students who are interested in beginning research in soft matter systems. To that end, a computational code is included that students can use right away to study complex fluids in equilibrium.

---

[*] Corresponding author. Electronic mail: ricardo.lopez@correo.fisica.uson.mx

# 1. Introduction

The present day importance of colloidal dispersions arises in large part from the impact of their industrial applications in contemporary societies. They are important also in the academic context because their stability is the result of the delicate balance of attractive and repulsive forces [1], which gives rise to complex phenomena of many-body interactions that cannot be fully understood using *ab initio* theories except for some simplified models. Hence, alternatives have been proposed based on effective interactions, which are solved for many particles using computers [2]. In addition to their complex equilibrium properties, colloidal dispersions represent a challenge in basic and applied research due to their rheological properties, such as viscosity. In colloidal dispersions, the viscosity can be significantly affected by the presence of polymer chains, as shall be discussed here.

In this article, we discuss the connection between fundamental molecular interactions and their manifestation in a macroscopically measured property such as the viscosity of a fluid, which is modeled and solved through computer simulations. In particular, we comment on how electrostatic and non – electrostatic interactions give rise to thermodynamic stability for systems in equilibrium, i.e. those that are not under the influence of external agents such as flow. We then discuss the competition of hydrodynamic and Brownian forces in colloidal dispersions in equilibrium, followed by the rheological characteristics of fluids. Finally, we present original results on the viscosity of a model complex fluid made up of two surfaces covered with polymers under steady shear. For such system, we find highly non – linear behavior of the friction coefficient, as an example of how macroscopic, measurable rheological properties of complex fluids can be accurately predicted starting from basic molecular interactions. The purpose of this work is, on the one hand, to serve as brief introduction to the rapidly growing field of soft matter physics. On the other, we place emphasis on the connection between basic molecular interactions and macroscopic, measurable properties using computer simulations as the solving tool. Lastly, we present original results on the prediction of the viscosity and friction of polymer brushes as an example of the usefulness of simulations as research tools in soft matter physics. Our intended audience are final – year undergraduate students and graduate students who are beginning research in soft condensed matter systems. We provide also a mesoscopic scale computational code; see Supplementary Information (S.I.), so that students can gain readily hands – on experience on these topics.

## 2. Basic interactions in colloidal dispersions

There are fundamental interactions that act between particles in colloidal dispersions, to which the presence of the solvent must be added. One of those is the so-called van der Waals (vdW) interaction, which is always present because it is the result of all fluctuations between induced electric dipoles surrounding each particle. When all interactions between two colloidal particles generated by induced dipoles are integrated, an expression for the total force between them is obtained, which is always attractive [1]. This force is a function of a single parameter, $A$, called the Hamaker constant, which depends on the nature of the solute and solvent and has units of energy. For example, for polystyrene in water at 25°C, $A = 10^{-20}$ J [3]. The explicit form of the vdW interaction can be obtained from the geometry of the colloidal particles. For short separations ($r$) between spherical particles (of radius $R$), with respect to their size, the surfaces can be considered flat and the vdW interaction is written as [1]:

$$U_{vdW}(r) = -\frac{A}{12\pi}\left(\frac{R}{r}\right)^2. \tag{1}$$

The attractive interaction shown in equation (1) is of short range, compared with the electrostatic interaction. However, when the distance between the colloids is small, the vdW interaction is extremely attractive, which may lead to the irreversible coagulation of colloidal particles. If the interaction in equation (1) was the only one present in a colloidal dispersion, one would expect that all particles coagulate eventually due to their attraction, and precipitate forming a mass of solid material. This clearly does not happen and it is partly due to other forces, which compensate to a greater or lesser extent the attractive vdW interaction.

The mechanisms that induce the presence of electric charges on the surface of colloidal particles in aqueous media are ionization or dissociation of functional groups on the surface, or by adsorption of ions with positive or negative charge [1]. For small surface electrical potential ($\psi_0$), the electrostatic potential energy between identical particles in a medium can be approximated as [1]:

$$U_E(r) = 4\pi\varepsilon \frac{a}{2+r/a} \psi_0^2 exp(-\kappa r). \qquad (2)$$

In equation (2), $a$ is the particle's diameter and $\kappa$ the Debye - Hückel constant, which is given by:

$$\kappa = \left(\frac{e^2 \sum n_{oi} z_i^2}{\varepsilon k_B T}\right)^2, \qquad (3)$$

where $e$ is the elementary electron charge, $n_{oi}$ is the concentration of ions of type $i$ away from the colloidal particle, $z_i$ is the valence of the ions of type $i$, $\varepsilon$ is the dielectric constant of the medium, $k_B$ is the Boltzmann constant and $T$ the absolute temperature. Equation (2) is not the "bare" Coulomb interaction, which decays as the inverse of the separation distance between the centers of mass of the particles. However, when there is a material medium in which charges are suspended, the pure Coulomb interaction is "dressed up" by the presence of the medium, which leads to the partial screening of the individual charges. Such screening gives rise to the exponential decay in equation (2), with the characteristic decay length (Debye length) given by the inverse of the expression in equation (3). The simultaneous presence of the interactions shown in equations (1) and (2), gives rise to an interaction potential function that develops a barrier, which prevents the coagulation of the particles. This potential is called DLVO, after the initials of its proponents (Derjaguin, Landau, Verwey and Overbeek) [1], and it is shown in Fig. 1 (red line), where the potential barrier is indicated by $E_{max}$.

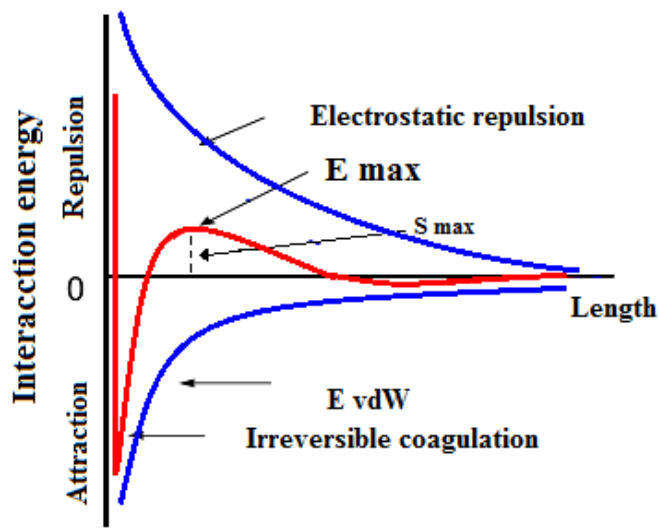

**Fig. 1.** DLVO interaction potential (red line). The *x* axis represents the distance separating the centers of mass of the colloidal particles while the *y* axis, the interaction resulting of particles under the influence of van der Waals (vdW) attraction and electrostatic repulsion. A maximum barrier of height Emax is obtained at relative distance Smax when the vdW and electrostatic interactions are added. Adapted from reference [3].

Two important aspects are of notice in Fig. 1. When the distances between the particles are too short, there is a deep minimum caused by the vdW interaction, equation (1), which will cause the particles to form aggregates and precipitate. Irreversible coagulation occurs due to the deep minimum in the energy, that is, the particles cannot be separated by Brownian collisions, for example. The dispersion will be in thermodynamic equilibrium, as it will be at the absolute minimum energy. The second aspect of importance shown in Fig. 1 is the presence of a potential barrier with value $E_{max}$, located at the relative distance $S_{max}$. When this barrier is considerably higher than the average thermal energy of the particles, it becomes very unlikely that the particles can coagulate after a collision and thus the dispersion will be kinetically stable. This type of stability, which is operative whenever there is a positive barrier in the potential energy function, may be enough to keep colloids dispersed in the medium, i. e., well separated from one another, but it is different from the thermodynamically stable state. The latter is the state of absolute minimum energy, which is reached of course when the relative distance between colloidal particles is close to zero, indicated as the irreversible coagulation point in Fig.1. Note that, given enough time, all the particles of the dispersion must reach the minimum energy at smaller distances like in Fig. 1. Thus, all colloidal dispersions will tend to coagulate, although the height of the barrier may be large enough to allow the dispersion to remain stable so that the particles can be kept apart during a long enough period of time for the use of the dispersion in practical purposes. In fact, there are instances where these two types of stability (kinetic and thermodynamic) compete in a system, yielding markedly different phases; see for example [4]. One mechanism known to prevent colloidal coagulation is the addition of polymers to the dispersion. When colloidal particles are coated with a polymer layer, which can occur by physical adsorption or when the polymers are anchored by means of chemical bonds, a repulsive force arises at short distance between them, which is of entropic origin [3].

## 3. Brownian motion and hydrodynamic interactions in colloidal dispersions

When an external force acts on the dispersion, the forces discussed in the previous section are not necessarily predominant, and hydrodynamic forces as well as Brownian collisions (at finite temperature) need to be taken into account. A hydrodynamic force is one that appears in a colloidal dispersion because of the flow of solvent, which in turn causes a change in the dynamics of the particles dissolved therein. One can calculate the hydrodynamic force acting on the surface of a particle of radius $a$ in a fluid if the following assumptions are made: (a) the flow is steady, so that its velocity $v_o$ is constant; (b) the viscous forces dominate over the inertial ones; (c) there are no external forces other than that causing the flow. In that case, the force on the colloidal particle is given by [5]:

$$F_H = 6\pi \eta a v_0 . \qquad (4)$$

The *intrinsic* viscosity of the fluid, considered here as a continuous medium where the particle is embedded, is represented by $\eta$. If the hydrodynamic force is known, we can use equation (4) to determine the particle size, which is useful in the paint industry, for example [6].

Brownian motion is characterized by erratic displacements driven by the temperature of the fluid and results from collisions between the molecules of the solvent, which is no longer considered a continuous medium. Solvent particles collide with the colloidal particle and transfer momentum to it, more or less chaotically due to the temperature at which the system is subjected. Hence, with increasing temperature Brownian collisions also increase. The characteristic Brownian force acting on a colloidal particle at a certain temperature can be written as

$$F_B = \frac{k_B T}{a} . \qquad (5)$$

Although Brownian motion is erratic, it can be quantified. The mean square displacement of a particle due to Brownian motion is equal to the number of collisions or "steps", $N$, that the particles experience multiplied by the size of the particle. Brownian forces are important for small particles, of the order of, or less than 1 μm. Solving accurately the motion a system with a large number of particles that interact via equations (1) and (2), in addition to the non – equilibrium interactions shown in equations (4) and (5) is virtually impossible, unless highly simplifying assumptions are made. Fortunately, an alternative approach can be used which takes advantage of the power of computers to carry out many operations with high precision in extremely short time. Such alternative is known generally as computer simulation [2], and can be applied to solve complex problems, such as the viscosity of colloidal dispersions, as is discussed in the following section.

## 4. Modeling the viscosity of colloidal dispersions with coarse grained computer simulations

We present briefly in this section some recent trends on the modeling of colloidal dispersions with soft particles. The central idea of molecular dynamics simulations [2] is relatively simple to state, see Fig. 2: a simulation box is constructed with a certain number of particles in it representing the molecules of the complex fluid one is interested in studying. Then, the equation of motion governing the particles' behavior must be solved for discreet increments in time ($\delta t$ in Fig. 2), which is usually Newton's second law of motion with a particular choice of interaction potential. The property of interest (represented by $A$ in Fig. 2, which can be energy, pressure, density, etc.) is then calculated for each configuration at the chosen time intervals. Averaging over long periods of time leads to accurate predictions of thermodynamic and structural properties. Typically, periodic boundary conditions [2] are applied to all the sides of the simulation box shown in Fig. 2, which means that one effectively can simulate a boundless, macroscopic fluid. However, as shown in the S.I., care must be exercised not to use too small a simulation box, as this can introduce artifacts into the simulations.

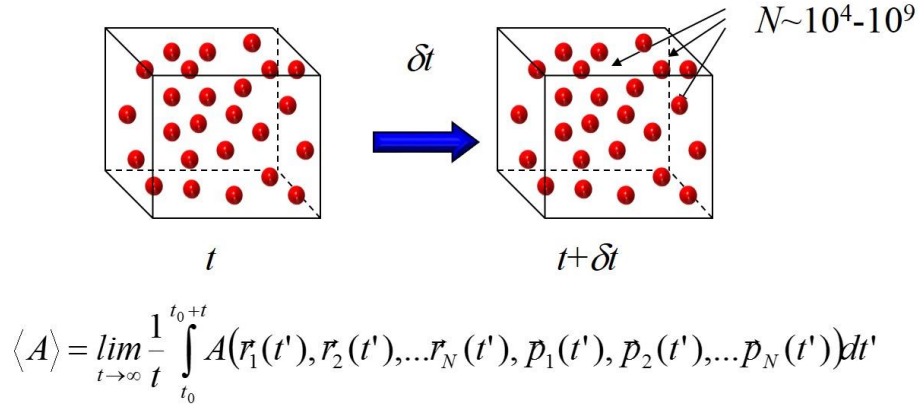

$$\langle A \rangle = \lim_{t \to \infty} \frac{1}{t} \int_{t_0}^{t_0+t} A(r_1(t'), r_2(t'), ... r_N(t'), p_1(t'), p_2(t'), ... p_N(t')) dt'$$

**Fig. 2.** The basic idea of molecular dynamics simulation. The equation of motion is integrated over discreet time intervals ($\delta t$), with the property of interest ($A$) being calculated as a function of the positions ($r_i$) and momenta ($p_i$) of all $N$ particles in the simulation box (shown here in red).

There are very sophisticated integration algorithms that solve the equation of motion of the particles in an essentially exact manner [2], which make of simulations a very attractive research tool. They have also the advantage that the user can choose the interaction model, and have total control over all its parameters. Another popular computer simulation technique is the one known as Monte Carlo (MC) [7], where the particles' configurations are not obtained from the integration of the equation of motion, as in molecular dynamics, but from a random choice of the position of the particles that make up the fluid of interest, see Fig. 3. MC simulations have the advantage that the interactions between particles do not have to be integrated because the configurations of particles are chosen according to a given algorithm, as schematically shown in Fig. 3. In MC simulations, once a final configuration is chosen, the property of interest (*A*) can be accurately calculated as an average over the ensemble, rather than over time, in contrast with molecular dynamics.

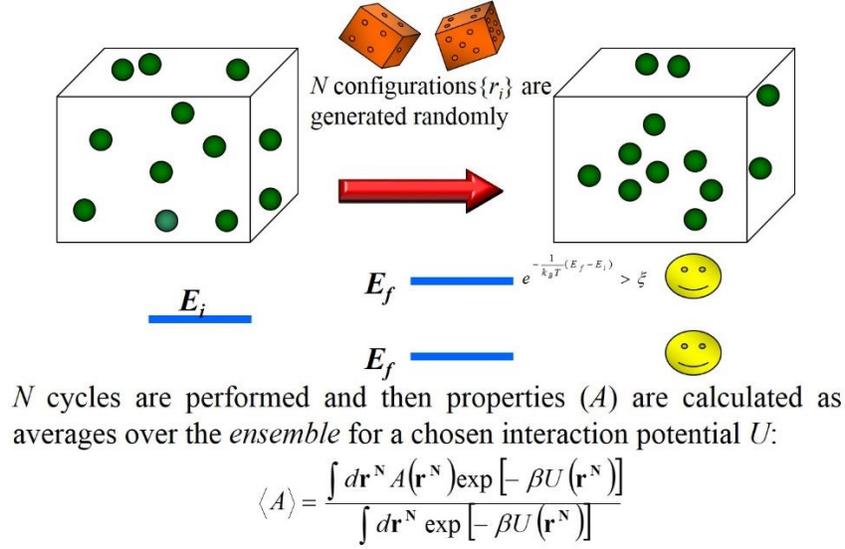

**Fig. 3.** Philosophy of Monte Carlo simulation. A spatial configuration of the particles in the simulation box (green disks) is chosen randomly (hence the dice) under the influence of an interaction model *U*, then a comparison is made between the energies of the initial configuration ($E_i$) and the final one ($E_f$). After that, an algorithm is applied that chooses between the initial and final configurations; in the case shown in the figure that is the so-called Metropolis algorithm [7]. Finally, the value of the property of interest (*A*) is obtained as an average over the ensemble of system.

One of the most successful computer simulation tools used presently is the technique known as "dissipative particle dynamics", or DPD [8, 9]. The main distinction between DPD and microscopic molecular dynamics simulations [2] is that the force acting between any two particles *i* and *j* in DPD is given not only by a conservative force ($\boldsymbol{F}_{ij}^{C}$), but also by dissipative ($\boldsymbol{F}_{ij}^{D}$), and random ($\boldsymbol{F}_{ij}^{R}$) forces. The total force acting on any given pair of particles is the sum of these three forces:

$$\boldsymbol{F}_{ij} = \sum_{i \neq j}^{N} [\boldsymbol{F}_{ij}^{C} + \boldsymbol{F}_{ij}^{D} + \boldsymbol{F}_{ij}^{R}]. \tag{6}$$

The explicit expressions for these forces, along with full details of the DPD model can be found in the S.I. It should be appreciated that the DPD model incorporates all the basic interactions reviewed in the previous sections; in particular, the van der Waals interaction, see equation (1), is effectively modeled by the conservative force, ($F_{ij}^C$). The Brownian interaction, equation (5), is represented in DPD by the random force ($F_{ij}^R$), while the hydrodynamic interactions (see equation (4)) are modeled in DPD by the viscous force, ($F_{ij}^D$). The entropic interactions that arise from the excluded volume effect, accompanied by the harmonic bonds used construct polymer chains as beads joined by springs, are incorporated also into ($F_{ij}^C$). Only the electrostatic interaction is missing from the DPD model, but it can readily be incorporated also [10]. An example of a working DPD code is provided in the S.I., along with the necessary tools to run it for complex fluids. It calculates structural and thermodynamic properties that are helpful for the basic understanding of fluids in equilibrium.

As Fig. 4 illustrates, among the advantages of DPD is that it does not solve the motion of particles at the atomic level, but rather at the mesoscopic level [11-14], which is particularly useful to reach scales in computer simulations that are comparable with those of the experiments.

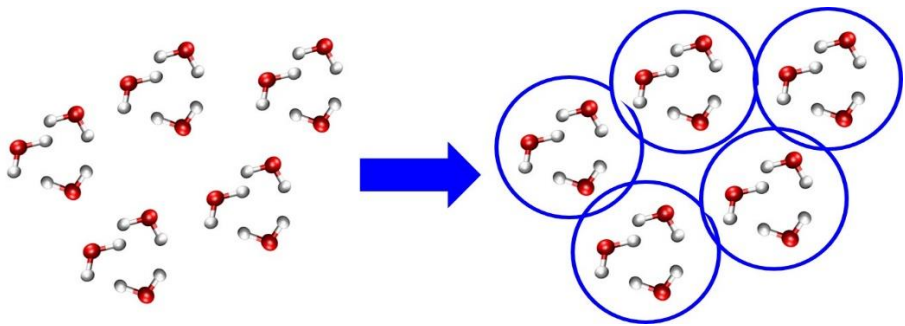

**Fig. 4.** Schematic illustration of the coarse – graining procedure in DPD. The particular case shown in this figure corresponds to a coarse – graining degree equal to 3, i. e., there are three water molecules grouped into a single DPD bead. The equation of motion is solved for the DPD particles (in blue).

The success of DPD simulations in equilibrium has driven the need to explore its applicability to non – equilibrium situations. As a case study, we report here original results on non-equilibrium DPD simulations of linear polymer chains grafted to a surface, forming "polymer brushes" under flow as shown in Fig. 5.

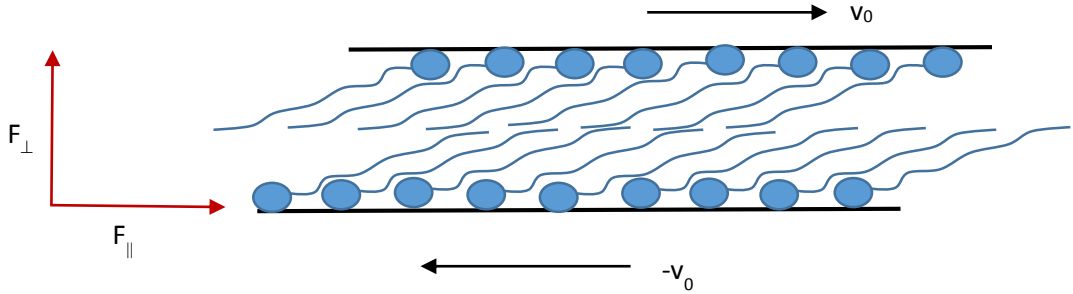

**Fig. 5**. Schematic illustration of the simulation setup used for non – equilibrium simulations. The circles in blue represent the beads grafted to the surface, while the lines represent the rest of the beads that make up the linear polymer chains. The diagram on the left defines the nomenclature for the forces perpendicular and parallel to the surface.

The motivation for this study is twofold. On the one hand, there are numerous industrial applications where it is necessary that the friction between surfaces be as small as possible so that films can easily slide past each other. On the other, much remains to be understood from the point of view of basic science concerning the mechanisms that give rise to low friction coefficients and/or low viscosity values. To predict values of the friction coefficient and viscosity one needs to perform non – equilibrium computer simulations where a steady external flow is applied to a confined fluid, as illustrated in Fig. 5. For the calculation of the friction coefficient ($\mu$) we used the equation $\mu = \langle F_x(\dot{\gamma}) \rangle / \langle F_z(\dot{\gamma}) \rangle$, see, for example, reference [15], where $F_x(\dot{\gamma})$ represents the magnitude of the force on the particles grafted onto each surface along the direction of the shear rate, $\dot{\gamma}$, and $F_z(\dot{\gamma})$ is the magnitude of the force on the particles, acting perpendicularly to the surfaces. The brackets indicate the time average of the forces. The viscosity ($\eta$) is obtained from the relation $\eta = \frac{\langle F_x(\dot{\gamma})\rangle/A}{\dot{\gamma}}$, where $A$ is the transversal area of the surface where polymers are grafted. The shear rate $\dot{\gamma}$ is equal to $2v_0^*/D^*$, where $v_0^*$ is the flow velocity exerted on the grafted monomers (see Fig. 5), and $D^*$ is the separation between the surfaces. The factor of 2 in the shear rate arises because both surfaces are moving, rather than only one. Using this model of polymer brushes under the influence of an external flow (Couette flow) has been shown to lead to the correct prediction of scaling exponents, among other phenomena [16, 17]. Full details of the simulation are omitted here for brevity but they can be found in the S.I., along with all the necessary information to reproduce them. The results of these simulations are presented in Fig. 6.

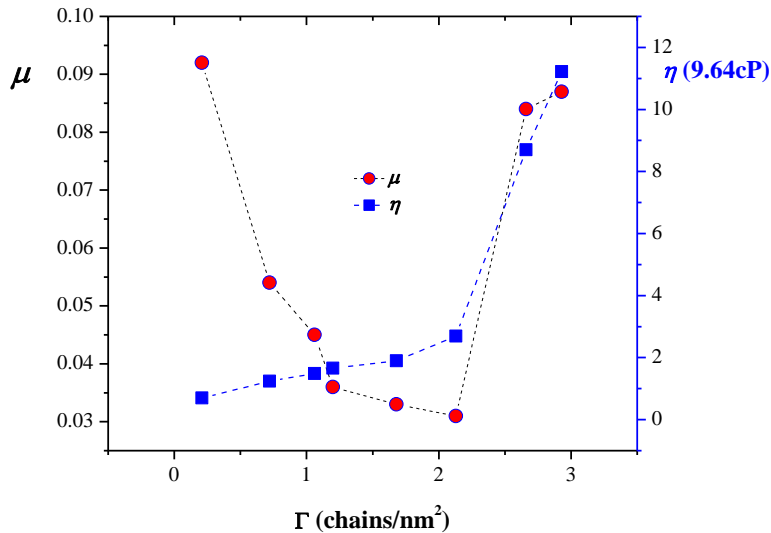

**Fig. 6**. Friction coefficient, $\mu$, of polymer brushes as a function of the number of polymer chains per nm$^2$ of substrate (filled circles). The viscosity ($\eta$) is shown also, see the filled squares. The lines are only guides for the eye. Notice the contrasting, non − monotonic behavior of the friction coefficient with respect to that of the viscosity.

As the number of chains per unit area ($\Gamma$) is increased, the friction coefficient displays a rapid decrease, reaching values as low as $\mu = 0.03$ when there are about 2 chains/nm$^2$, see Fig. 6.

The viscosity increases monotonically as $\Gamma$ is increased, in sharp contrast with the behavior of $\mu$. This unexpected behavior of these two rheological properties has its origin in the increasing osmotic pressure when $\Gamma$ is increased, which translates as an increase in $\langle F_z \rangle$ that dominates over the increase in $\langle F_x \rangle$, therefore $\mu$ decreases. Nonetheless $\langle F_x \rangle$ does grow with $\Gamma$, and that fact is responsible for the monotonic increase in the viscosity, $\eta$. As more chains are grafted on the surfaces the number of solvent particles must be reduced so that the overall density remains constant. This means that the layer of fluid between the opposing polymer brushes is increasingly thinner, leading to relatively large values of both $\mu$ and $\eta$, when $\Gamma$ is close to 3 chains/nm$^2$, which incidentally is close to the maximum grafting density that can be reached with the experimental techniques presently available [18]. The conclusion of these simulations is that the optimal value of the grafting density of the polymer chains we have modeled is about 2 chains/nm$^2$, because in that case the friction coefficient is at its minimum, while the viscosity has only modestly increased. Moreover, these simulations allow us to predict macroscopic rheology properties based on models of basic molecular interactions, thereby underlying the potential of these modern computational tools in the study of colloidal dispersions.

## 6. Conclusions

In this work we have reviewed the basic interactions of colloid dispersions as well as some of their rheological properties. We emphasized that much can be learned using relatively simple models and that it is also possible to obtain quantitative trends using innovative techniques such as computational simulation. Our purpose here was to show how macroscopic properties of complex fluids, such as the viscosity, can be accurately obtained from basic molecular interactions, solved for a large number of particles using appropriately adapted algorithms. It is our hope that this work can be useful as an introduction to carry out research in this vibrant field.

## Acknowledgements

AGG acknowledges C. Ávila, C. Pastorino, Z. Quiñones and E. Rivera for useful input. AGG, MABA and RLE thank IFUASLP for its hospitality.

**Supplementary Information**

In the Supplementary Information that accompanies this article, we have included a simplified but working version of a DPD code that both students and researchers new to this field can use to carry out computer simulations of model complex fluids, such as polymers in solution. The code calculates several structural and thermodynamic properties, such as pair distribution functions, density profiles, pressure, and interfacial tension.